\documentclass[aps,preprint]{revtex4}%
\usepackage{amsfonts}
\usepackage{amsmath}
\usepackage{amssymb}
\usepackage{graphicx}%
\providecommand{\U}[1]{\protect\rule{.1in}{.1in}}

\begin{document}
\preprint{ }
\title[1/N ]{Spectral function in QED$_{3}$}
\author{Yuichi Hoshino}
\affiliation{Kushiro National College of Technology,Otanoshike Nishi
-32-1,Kushiro,Hokkaido, 084-0916,Japan}
\affiliation{Centor for the Subatomic Structure of Matter(CSSM),University of Adelaide,AUSTRALIA5005}
\author{}
\affiliation{}
\author{}
\affiliation{}

\begin{abstract}
We discuss the structure of the dressed fermion propagator in unquenched QED3
based on spectral function of photon.In this approximation infrared
divergences that appeared in quenched case turns out to be soft.The dimension
full coupling constant naturally appears as an infrared mass scale in this
case.We find the reliable results for the effects of vacuum polarization for
the dressed fermion propagator.The lowest order fermion spectral function has
logarithmically divergent Coulomb energy as well as self-energy,whch plays the
role of confinement and dynamical mass generation.In our model finiteness
condition of vacuum expectation value is equivalent to choose the scale of
physical mass which is expected in the $1/N$ approximation.

\end{abstract}
\volumeyear{year}
\volumenumber{number}
\issuenumber{number}
\eid{identifier}
\date[Date text]{date}
\received[Received text]{date}

\revised[Revised text]{date}

\accepted[Accepted text]{date}

\published[Published text]{date}

\startpage{101}
\endpage{102}
\maketitle
\tableofcontents

\section{ \qquad Introduction}

About 25 years ago it was pointed out that high temprerature limit of the
field theory is descrived by the same theory with less-dimension and it
suffers from severe infrared divergences with dimensionful coupling
constant[1]. Before these argument infrared divergences associated with
massless particle such as photon and graviton has been discussed to determine
the infrared structure of one particle state[2].These are known as cut
structure or infrared behaviour of the propagator near mass shell.In this kind
of works renormalization group analysis or method of spectral function had
been applied[2].However in the analysis of three dimensional theory these
approaches have not been used.The main reason may be in the dimensionfull
coupling constant and the super renormalizability of the model.But in[2,3,4]
infrared divergence near the mass shell has been given in a model independent
way.Of course the topological mass soften infrared divergence but it is
limited to parity violating case and we concentrate ourselves to parity
conserving case[5].At the same time (2+1) dimensional QED with massless
fermion has been shown its dynamical chiral symmetry breaking by the analysis
of Dyson-Schwinger (D-S) equation and Lattice simulation [6,7,8,9,10,11,12]and
the massless pair instability has been shown by solving Behte-Salpeter (B-S)
equation with $1/N$ approximation[13].In condensed matter physics
(2+1)-dimensional fermionic system shows super-conductivity phase. It has been
pointed out that the d-wave superconductor-insulator transition at $T=0$ is
analougas to the dynamical mass generation in QED$_{3}$.And QED$_{3}$ is
referred to as an effective theory of phase transition for its pairing
instability of massless fermion[14,15].In the previous work we studied fermion
propagator based on the mass singularity in QED$_{3}$ in quenched
approximation with finite bare photon mass in spectral representation[3].After
exponentiation one photon matrix element we have the explicit form of the
propagator with confining properties,dynamical mass generation as we got in
the D-S equation.The purpose of this paper is to improve quenched
approximation by taking into account of vacuum polarization.Other approaches
has shown that the screening effects soften the infrared behaviour
[1,4,6].Especially in [1] order $e^{4\text{ }}\ln(e^{2})$ self-energy of
massless fermion was improved by spectral function of photon with vacuum
polarization of massless fermion.In this work we use the same spectral
function of dressed photon propagator and get the non-perturbative effects by
integrating the quenched spectral function of fermion with bare photon mass.As
a result linear infrared divergences turned out to be logarithmic one,and
$\ln(\mu\left\vert x\right\vert )$ is converted to $\ln(e^{2}\left\vert
x\right\vert )$ in the spectral function.Thus we can remove an infrared
cut-off from the dressed fermion propagator away from the threshold, where the
coupling constant $e^{2}$ plays the role of photon mass.In comparison our
results with the analysis by unquenched Dyson-Schwinger equation,our solution
is consistent with the latter excepts for wave renormalization $Z_{2}^{-1}%
=0.$In section II spectral representation of the fermion propagator are
defined and we show the way to determine it based on LSZ reduction formula and
low-energy theorem.In section III we evaluate the full propagator for quenched
case with photon mass as an infrarfed cut-off and improve it by the photon
spectral function for unquenched case.In Section IV is devoted to the analysis
in momentum space of these solutions and the comparison with Dyson-Schwinger analysis[9,11,12].

\section{Calculating the spectraly weighted propagator}

\subsection{\bigskip Definition of the spectral function}

In this section we show how to evalute the fermion propagator non
pertubatively by the spectral represntation which preserves unitarity and
analyticity[2,3,4].The spectral function of the fermion is defined
\begin{align}
S_{F}(s^{\prime})  &  =P\int ds\frac{\gamma\cdot p\rho_{1}(s)+\rho_{2}%
(s)}{s^{\prime}-s}+i\pi(\gamma\cdot p\rho_{1}(s^{\prime})+\rho_{2}(s^{\prime
})),\\
\rho(p)  &  =\frac{1}{\pi}\operatorname{Im}S_{F}(p)=\gamma\cdot p\rho
_{1}(p)+\rho_{2}(p)\nonumber\\
&  =(2\pi)^{2}\sum_{n}\delta(p-p_{n})\left\langle 0|\psi(x)|n\right\rangle
\left\langle n|\overline{\psi}(0)|0\right\rangle .
\end{align}
In the quenched approximation the state $|n>$ stands for a fermion and
arbitrary numbers of photons,%
\begin{equation}
|n>=|r;k_{1},...,k_{n}>,r^{2}=m^{2},
\end{equation}
we have the solution for the spectral function $\rho(p)$ which is written
symbolically%
\begin{align}
\rho(p)  &  =\int\frac{md^{2}r}{r^{0}}\sum_{n=0}^{\infty}\frac{1}{n!}\left(
\int\frac{d^{3}k}{(2\pi)^{2}}\theta(k^{0})\delta(k^{2})\sum_{\epsilon}\right)
_{n}\delta(p-r-\sum_{i=1}^{n}k_{i})\nonumber\\
&  \times\left\langle \Omega|\psi(x)|r;k_{1},...,k_{n}\right\rangle
\left\langle r;k_{1},...,k_{n}|\overline{\psi}(0)|\Omega\right\rangle .
\end{align}
Here the notations
\[
(f(k))_{0}=1,(f(k))_{n}=%
{\displaystyle\prod\limits_{i=1}^{n}}
f(k_{i})
\]
have been introduced to show the phase space of each photons.To evaluate the
contribution of soft photons,first we consider the situation when only the
$n$th photon is soft.We define the matrix element
\begin{equation}
T_{n}=\left\langle \Omega|\psi|r;k_{1},....,k_{n}\right\rangle .
\end{equation}
We consider $T_{n}$ for $k_{n}^{2}\neq0,$continue off the photon mass shell by
LSZ reduction formula:%
\begin{align}
T_{n}  &  =\epsilon_{n}^{\mu}T_{n\mu},\nonumber\\
\epsilon_{n}^{\mu}T_{n}^{\mu}  &  =\frac{i}{\sqrt{Z_{3}}}\int d^{3}%
y\exp(ik_{n}\cdot y)\square_{y}\left\langle \Omega|T\psi(x)\epsilon^{\mu
}A_{\mu}(y)|r;k_{1},...,k_{n-1}\right\rangle \nonumber\\
&  =-\frac{i}{\sqrt{Z_{3}}}\int d^{3}y\exp(ik_{n}\cdot y)\left\langle
\Omega|T\psi(x)\epsilon^{\mu}j_{\mu}(y)|r;k_{1},...,k_{n-1}\right\rangle ,
\end{align}
provided%
\begin{align}
\square_{x}T(\psi A_{\mu}(x)  &  =T\psi\square_{x}A_{\mu}(x)=T\psi(-j_{\mu
}(x)+\frac{d-1}{d}\partial_{\mu}^{x}(\partial\cdot A(x)),\\
\partial\cdot A^{(+)}|phys  &  >=0,
\end{align}
where $d$ is a gauge fixing parameter.$T_{n}$ satisfies
Ward-Takahashi-identity%
\begin{equation}
k_{n\mu}T_{\mu}^{n}(r;k_{1},...,k_{n})=e\exp(ik_{n}\cdot x)\left\langle
\Omega|\psi(x)|r;k_{1},....,k_{n-1}\right\rangle ,
\end{equation}
provided%
\begin{equation}
\partial_{\mu}^{y}T(\psi(x)J_{\mu}(y))=-e\psi(x)\delta(x-y).
\end{equation}
By translational invariance%
\begin{equation}
\psi(x)=\exp(-ip\cdot x)\psi(0)\exp(ip\cdot x),
\end{equation}
we get the usual form%
\begin{equation}
k_{n\mu}T_{\mu}^{n}(r;k_{1},...k_{n})=eT_{n-1}(r;k_{1},...k_{n-1}),r^{2}%
=m^{2}.
\end{equation}
By the low-energy theorem the fermion pole term in $T_{\mu}^{n}$ is dominant
for the infrared singularity in $k_{n}^{\mu}$.Inclusion of regular terms and
their contribution to Ward-identities are given for the scalar
case[2].Hereafter we consider the leading pole term for simplicity.

\subsection{\bigskip Approximation to the spectral function}

One-photon matrix element $T_{1}$ which is given in [3,16]%
\begin{align}
&  T_{1}=\left\langle \Omega|\psi(x)|r;k\right\rangle =\left\langle
in|T(\psi_{in}(x),ie\int d^{3}y\overline{\psi}_{in}(y)\gamma_{\mu}\psi
_{in}(y)A_{in}^{\mu}(y)|r;k\text{ }in\right\rangle \nonumber\\
&  =ie\int d^{3}yd^{3}zS_{F}(x-z)\gamma_{\mu}\delta^{(3)}(y-z)\exp(i(k\cdot
y+r\cdot z))\epsilon^{\mu}(k,\lambda)U(r,s)\nonumber\\
&  =-ie\frac{1}{(r+k)\cdot\gamma-m+i\epsilon}\gamma_{\mu}\epsilon^{\mu
}(k,\lambda)\exp(i(r+k)\cdot x)U(r,s).
\end{align}
Here $U(r,s)$ is a four component free particle spinor with \ positive
energy.If we sum infinite numbers of photon in the final state as in
(4),assuming pole dominance for $k_{n}^{\mu}$ we have a simplest solution to
$T_{n}$ in (12)
\begin{align}
T_{n}|_{k_{n}^{2}=0}  &  =T_{1}T_{n-1}\nonumber\\
&  =e\frac{\gamma\cdot\epsilon_{n}}{\gamma\cdot(r+k_{n})-m}T_{n-1}.
\end{align}
From this relation we obtain the $n$-photon matrix element $T_{n}$ as the
direct products of $T_{1}$%
\begin{equation}
\left\langle \Omega|\psi(x)|r;k_{1},...,k_{n}\right\rangle \left\langle
r;k_{1},..k_{n},\overline{\psi}(0)|\Omega\right\rangle \rightarrow%
{\displaystyle\prod\limits_{j=1}^{n}}
T_{1}(k_{j})\overline{T}_{1}(k_{j}).
\end{equation}
In this way we have an approximate solution of (4) by exponentiation of the
one-photon matrix element
\begin{equation}
\overline{\rho}(x)=-(i\gamma\cdot\partial+m)\int\frac{d^{2}r}{(2\pi)^{2}r^{0}%
}\exp(ir\cdot x)\exp(F),
\end{equation}%
\begin{align}
F  &  =\sum_{one\text{ }photon}\left\langle \Omega|\psi(x)|r;k\right\rangle
\left\langle r;k|\overline{\psi}(0)|\Omega\right\rangle \nonumber\\
&  =\int\frac{d^{3}k}{(2\pi)^{2}}\delta(k^{2})\theta(k^{0})\exp(ik\cdot
x)\sum_{\lambda,s}T_{1}\overline{T_{1}}.
\end{align}%
\begin{align}
\sum_{\lambda,s}T_{1}\overline{T}_{1}  &  =\frac{(r+k)\cdot\gamma+m}%
{(r+k)^{2}-m^{2}}\gamma^{\mu}\frac{(\gamma\cdot r+m)}{2m}\frac{(r+k)\cdot
\gamma+m}{(r+k)^{2}-m^{2}}\gamma^{\nu}\Pi_{\mu\nu}\nonumber\\
&  =-e^{2}(\frac{\gamma\cdot r}{m}+1)[\frac{m^{2}}{(r\cdot k)^{2}}+\frac
{1}{(r\cdot k)}+\frac{d-1}{k^{2}}],\nonumber\\
r^{2}  &  =m^{2},k^{2}=0.
\end{align}
Here $\Pi_{\mu\nu}$ is the polarization sum%
\begin{equation}
\Pi_{\mu\nu}=\sum_{\lambda}\epsilon_{\mu}(k,\lambda)\epsilon_{\nu}%
(k,\lambda)=-g_{\mu\nu}-(d-1)\frac{k_{\mu}k_{\nu}}{k^{2}}.
\end{equation}
And the free photon propagator has the form%
\begin{equation}
D_{0}^{\mu\nu}=\frac{1}{k^{2}+i\epsilon}[g^{\mu\nu}-\frac{k^{\mu}k^{\nu}%
}{k^{2}}+d\frac{k^{\mu}k^{\nu}}{k^{2}}].
\end{equation}
The fermion propagator is written explicitly in the following form in quenched
case%
\begin{align}
S_{F}(x)  &  =-(i\gamma\cdot\partial+m)\frac{\exp(-m\left\vert x\right\vert
)}{4\pi\left\vert x\right\vert }\nonumber\\
&  \times\exp(-e^{2}\int\frac{d^{3}k}{(2\pi)^{2}}\exp(ik\cdot x)\theta
(k^{0})\delta(k^{2})[\frac{m^{2}}{(r\cdot k)^{2}}+\frac{1}{(r\cdot k)}%
+\frac{d-1}{k^{2}}]),\nonumber\\
\left\vert x\right\vert  &  =\sqrt{-x^{2}}.
\end{align}
Here $\delta(k^{2})$ is read as the imaginary part of the photon
propagator.For unquenched case we use the dressed photon with massless fermion
loop with $N$ \ flavours.Spectral functions for free and dressed photon are
given by [1,3,4,17]%
\begin{align}
\rho^{(0)}(k)  &  =\delta(k^{2}-\mu^{2}),\nonumber\\
\rho^{D}(k)  &  =\operatorname{Im}D_{F}(k)=\frac{c\sqrt{k^{2}}}{k^{2}%
(k^{2}+c^{2})},c=\frac{e^{2}N}{8}.
\end{align}
In this case one photon matrix element is modified to%
\begin{equation}
F=-e^{2}\int\frac{d^{3}k}{(2\pi)^{2}}\exp(ik\cdot x)[\operatorname{Im}%
D_{F}(k)[\frac{m^{2}}{(r\cdot k)^{2}}+\frac{1}{(r\cdot k)}-\frac{1}{k^{2}%
}]+\frac{d}{k^{4}}].
\end{equation}
To evaluate the $F$ we use the position space propagator in the next
section.The spectral representation for photon propagator is written
\begin{align}
iD_{F}(x)  &  =\frac{1}{\pi}\int\frac{d^{3}k}{(2\pi)^{3}}\exp(ik\cdot
x)\int_{0}^{\infty}dp^{2}\frac{\rho^{D}(p)}{k^{2}+p^{2}}\nonumber\\
&  =\int_{0}^{\infty}dp^{2}\frac{\exp(-\sqrt{p^{2}}\left\vert x\right\vert
)}{4\pi\left\vert x\right\vert }\rho^{D}(p^{2}),
\end{align}
as a linear superposition of the photon with $\sqrt{p^{2}}$ mass.

\section{Approximate solution in position space}

\subsection{Quenched case}

To evaluate the function $F$%
\begin{align}
F &  =-e^{2}\int\frac{d^{3}k}{(2\pi)^{2}}\exp(ik\cdot x)\theta(k^{0}%
)\delta(k^{2})[\frac{m^{2}}{(r\cdot k)^{2}}+\frac{1}{(r\cdot k)}+\frac
{d-1}{k^{2}}]\nonumber\\
&  =F_{1}+F_{2}+F_{L},
\end{align}
it is helpful to use the exponential cut-off(infrared cut-off)[2,3,18].In the
appendices the way to evaluate $F$ is given.By using the photon propagator
with bare mass $\mu$ we obtain%
\begin{equation}
F\simeq\frac{e^{2}(d-2)}{8\pi\mu}+\frac{\gamma e^{2}}{8\pi\sqrt{r^{2}}}%
+\frac{e^{2}}{8\pi\sqrt{r^{2}}}\ln(\mu\left\vert x\right\vert )-\frac{e^{2}%
}{8\pi}\left\vert x\right\vert \ln(\mu\left\vert x\right\vert )-\frac{e^{2}%
}{16\pi}\left\vert x\right\vert (d-3+2\gamma),
\end{equation}
where $\gamma$ is an Euler constant.In this case linear infrared divergence
may cancells by higher order correction or away from threshold.At present we
omitt them here with constant term.Linear term in $\left\vert x\right\vert $
is understood as the finite mass shift from the form of the propagator in
position space (29) and $\left\vert x\right\vert \ln(\mu\left\vert
x\right\vert )$ term is position dependent mass
\begin{align}
m &  =\left\vert m_{0}+\frac{e^{2}}{16\pi}(d-3+2\gamma)\right\vert ,\\
m(x) &  =m+\frac{e^{2}}{8\pi}\ln(\mu\left\vert x\right\vert ),r^{2}=m^{2},
\end{align}
which we will discuss in section IV.Here we see that the position space
propagator is written as free one with physical mass $m$ multiplied by quantum
correction as%
\begin{align}
\frac{\exp(-m\left\vert x\right\vert )}{4\pi\left\vert x\right\vert }\exp(F)
&  =\frac{\exp(-m\left\vert x\right\vert )}{4\pi\left\vert x\right\vert }%
(\mu\left\vert x\right\vert )^{D-C\left\vert x\right\vert },\nonumber\\
D &  =\frac{e^{2}}{8\pi m},C=\frac{e^{2}}{8\pi}.
\end{align}
From the above form we see that $D$ acts to change the power of $\left\vert
x\right\vert $ and plays the role of anomalous dimension of the
propagator[3,18].To search the stability of massless $e^{+}e^{-}$ composite in
the lattice simulation Colomb energy and self energy were
considered[7].Recently this problem has analysed in the B-S amplitude and
shows the pairing instability which signals dynamical rearrangement of the
vaccum[8].In our case Coulomb energy for two electrons in two dimension is%
\[
\rho(x)=e\delta^{(2)}(x-x(t)),\rho(y)=e\delta^{(2)}(y),
\]%
\begin{align}
-E_{C} &  =-\frac{e^{2}}{2}\int d^{2}xd^{2}y\rho(x)\frac{K_{0}(\mu\left\vert
x\right\vert )}{\pi}\rho(y)\nonumber\\
&  \rightarrow\frac{e^{2}}{2\pi}(\ln(\frac{\mu\left\vert x\right\vert }%
{2})+\gamma)+O(\mu).
\end{align}
This is qualitatively the same with $F_{2}$%
\[
F_{2}=-\frac{e^{2}}{8\pi}\operatorname{Ei}(\mu\left\vert x\right\vert
)\rightarrow\frac{e^{2}}{8\pi}(\gamma+\ln(\mu\left\vert x\right\vert
))+O(\mu).
\]
If we compare these result with ours$,-F$ is a sum of self energy of electron
and Coulomb energy of two electron $e^{-}e^{-}$ and it is short ranged and
positive at short distance $\mu\left\vert x\right\vert \leq1$ due to the
exponential cut-off .To see the differece between our approximation in three
and four dimension,here we show the result in four dimension[2].In
four-dimension the photon propagator is%
\begin{equation}
D_{F}^{(0)}(x)_{+}=\frac{\mu K_{1}(\mu\left\vert x\right\vert )}{4\pi
^{2}i\left\vert x\right\vert }=\frac{1}{4\pi^{2}ix^{2}}+O(\mu^{2}\ln
(\mu\left\vert x\right\vert ))+O(\mu^{4}).
\end{equation}
We obtain by the parameter integral
\begin{align}
F_{1} &  =-ie^{2}m^{2}\int_{0}^{\infty}\frac{\alpha d\alpha}{4\pi
^{2}i(x+\alpha r)^{2}}\exp(-\alpha\mu r)=\frac{e^{2}}{8\pi^{2}}\ln(\mu
^{2}\left\vert x\right\vert ^{2}),\\
F_{2} &  =-e^{2}\int_{0}^{\infty}\frac{d\alpha}{4\pi^{2}(x+\alpha r)^{2}%
}=-\frac{e^{2}}{4\pi^{2}\sqrt{r^{2}}\left\vert x\right\vert }=-\frac{e^{2}%
}{4\pi^{2}m\left\vert x\right\vert },\\
F_{L} &  =-i(d-1)e^{2}\frac{\partial}{\partial\mu^{2}}\frac{\mu K_{1}%
(\mu\left\vert x\right\vert )}{4\pi^{2}i\left\vert x\right\vert }%
=-\frac{(d-1)e^{2}}{16\pi^{2}}\ln(\mu^{2}\left\vert x\right\vert ^{2})+O(\mu).
\end{align}
To evaluate $F_{L}$, $\delta(k^{2})/k^{2}$ in (23) is replaced by
$-\delta^{^{\prime}}(k^{2})$ in the definition of $F.$The $-F_{2}$ is
interpreted as the Coulomb energy of $e^{-}e^{-}$ separated by $\left\vert
x\right\vert $ devided by $m$.$F_{2}$ is finite in the infrared and does not
contribute for the infrared singularity.Therefore the leading log correction
leads the well-known form by Fourier transformation%
\[
S_{F}(p)=-\int d^{4}x\exp(-ip\cdot x)(i\gamma\cdot\partial+m)\frac
{mK_{1}(m\left\vert x\right\vert )}{4\pi^{2}i\left\vert x\right\vert }(\mu
^{2}\left\vert x\right\vert ^{2})^{D}%
\]%
\begin{equation}
S_{F}(p)\simeq\frac{\gamma\cdot p+m}{m^{2}(1-p^{2}/m^{2})^{1-D}}%
,D=\frac{\alpha(d-3)}{2\pi},\alpha=\frac{e^{2}}{4\pi},
\end{equation}
near $p^{2}=m^{2}.$

\subsection{\bigskip Unquenched case}

Here we apply the spectral function of photon to evaluate the unquenched
fermion proagator.We simply integrate the function $F(x,\mu)$ for quenched
case which is given in (26),where $\mu$ is a photon mass.Spectral function of
photon is given in (22) in the Landau gauge $d=0$%
\begin{align}
\rho^{D}(\mu) &  =\frac{c}{\mu(\mu^{2}+c^{2})},\nonumber\\
Z_{3}^{-1} &  =\int_{0}^{\infty}2\rho(\mu)\mu d\mu=\pi.
\end{align}
An improved $F$ is written as dispersion integral%
\begin{align}
\widetilde{F} &  =\int_{0}^{\infty}2F(\mu)\rho^{D}(\mu)\mu d\mu\nonumber\\
&  =\frac{e^{2}}{8\pi c}(4)\ln(\frac{\mu}{c})+\frac{\gamma e^{2}}{8m}%
+\frac{e^{2}}{8m}\ln(c\left\vert x\right\vert )\nonumber\\
&  -\frac{e^{2}}{8}\left\vert x\right\vert \ln(c\left\vert x\right\vert
)-\frac{e^{2}}{16}\left\vert x\right\vert (3-2\gamma),
\end{align}
where linear divergent term is regularized by cut-off $\mu$%
\begin{equation}
\int_{\mu}^{\infty}\frac{2cd\mu}{\mu(\mu^{2}+c^{2})}=-\frac{2}{c}\ln(\frac
{\mu}{c}).
\end{equation}
In this way the linear infrared divergences turn out to be a logarithmic
divergence in the first term and $\mu$ in the other logarithms is convereted
by $c$ under the dispersion integral.Hereafter we neglect the $\ln(\mu/c)$
term in Euclid space.The fermion propagator with $N$ flavours in position
space is modified to
\begin{align}
S_{F}(x) &  =-(i\gamma\cdot\partial+m)\overline{\rho}(x)\nonumber\\
\overline{\rho}(x) &  =\frac{\exp(-m\left\vert x\right\vert )}{4\pi\left\vert
x\right\vert }\exp(\widetilde{F})\\
&  =\frac{\exp(-\left\vert m_{0}+B\right\vert \left\vert x\right\vert )}%
{4\pi\left\vert x\right\vert }(c\left\vert x\right\vert )^{D-C\left\vert
x\right\vert }(\frac{\mu}{c})^{\beta}\exp(\frac{\gamma e^{2}}{8m}),
\end{align}%
\begin{equation}
B=\frac{c}{2N}(3-2\gamma),\beta=\frac{4}{N\pi},C=\frac{c}{N},D=\frac{c}{Nm}.
\end{equation}
At large $N$ the function damps slowly with fixed $c,$where mass changing
effect is small.For small $N$ the function damps fast and the short distant
part is dominant for mass changing effect.

\section{\bigskip In momentum space}

Now we turn to the fermion propagator in momentum space.The momentum space
propagator is given by Fourier transform%
\begin{align}
S_{F}(p)  &  =\int d^{3}x\exp(-ip\cdot x)S_{F}(x)\nonumber\\
&  =-\int d^{3}x\exp(-ip\cdot x)(i\gamma\cdot\partial+m)\frac{\exp
(-m\left\vert x\right\vert )}{4\pi\left\vert x\right\vert }\exp(F(x)).
\end{align}
where we have
\begin{equation}
\exp(F(x))=A(c\left\vert x\right\vert )^{D-C\left\vert x\right\vert }%
,A=\exp(\frac{e^{2}\gamma}{8m}),D=\frac{c}{Nm},C=\frac{c}{N}.
\end{equation}
It is known that%
\begin{align}
&  \int d^{3}x\exp(-ip\cdot x)\frac{\exp(-m\left\vert x\right\vert )}%
{4\pi\left\vert x\right\vert }(c\left\vert x\right\vert )^{D}\nonumber\\
&  =c^{D}\frac{\Gamma(D+1)\sin((D+1)\arctan(\sqrt{-p^{2}}/m)}{\sqrt{-p^{2}%
}(p^{2}+m^{2})^{(D+1)/2}}\\
&  =\frac{1}{p^{2}+m^{2}}\text{ for }D=0,\nonumber\\
&  =\frac{2m}{(p^{2}+m^{2})^{2}}\text{ for }D=1.
\end{align}
for Euclidean momentum $p^{2}\leq0$.$S_{F}(p)$ in Minkowski space is given in
the Appendix B.To use the above formula for the case of mass generation
$C\neq0$ we use Laplace transform[3]
\begin{equation}
F(s)=\int_{0}^{\infty}d\left\vert x\right\vert \exp(-(s-m)\left\vert
x\right\vert )\left(  c\left\vert x\right\vert \right)  ^{-C\left\vert
x\right\vert }(s\geq0).
\end{equation}
This function shifts the mass and we get the propagator
\begin{equation}
S_{F}(p)=(\gamma\cdot p+m)Ac^{D}\Gamma(D+1)\int_{0}^{\infty}F(s)ds\frac
{\sin((D+1)\arctan(\sqrt{-p^{2}}/(m-s))}{\sqrt{-p^{2}}(p^{2}+(m-s)^{2}%
)^{(D+1)/2}}.
\end{equation}
At $D=0$ and $1$ we see
\begin{equation}
S_{F}(p)=(\gamma\cdot p+m)A\int_{0}^{\infty}F(s)ds\frac{1}{(p^{2}+(m-s)^{2}%
)},(D=0).
\end{equation}%
\begin{equation}
S_{F}(p)=(\gamma\cdot p+m)Ac\int_{0}^{\infty}F(s)ds\frac{2\left\vert
m-s\right\vert }{(p^{2}+(m-s)^{2})^{2}},(D=1).
\end{equation}
$N$ dependence of $m\overline{\rho}(x)$ and its Fourier transfom $m\rho(p)$
for $D=1$ are shown in Fig.1 and Fig.2.%

{\parbox[b]{3.2214in}{\begin{center}
\includegraphics[
trim=0.000000in 0.000000in 0.053341in 0.053341in,
height=3.2214in,
width=3.2214in
]%
{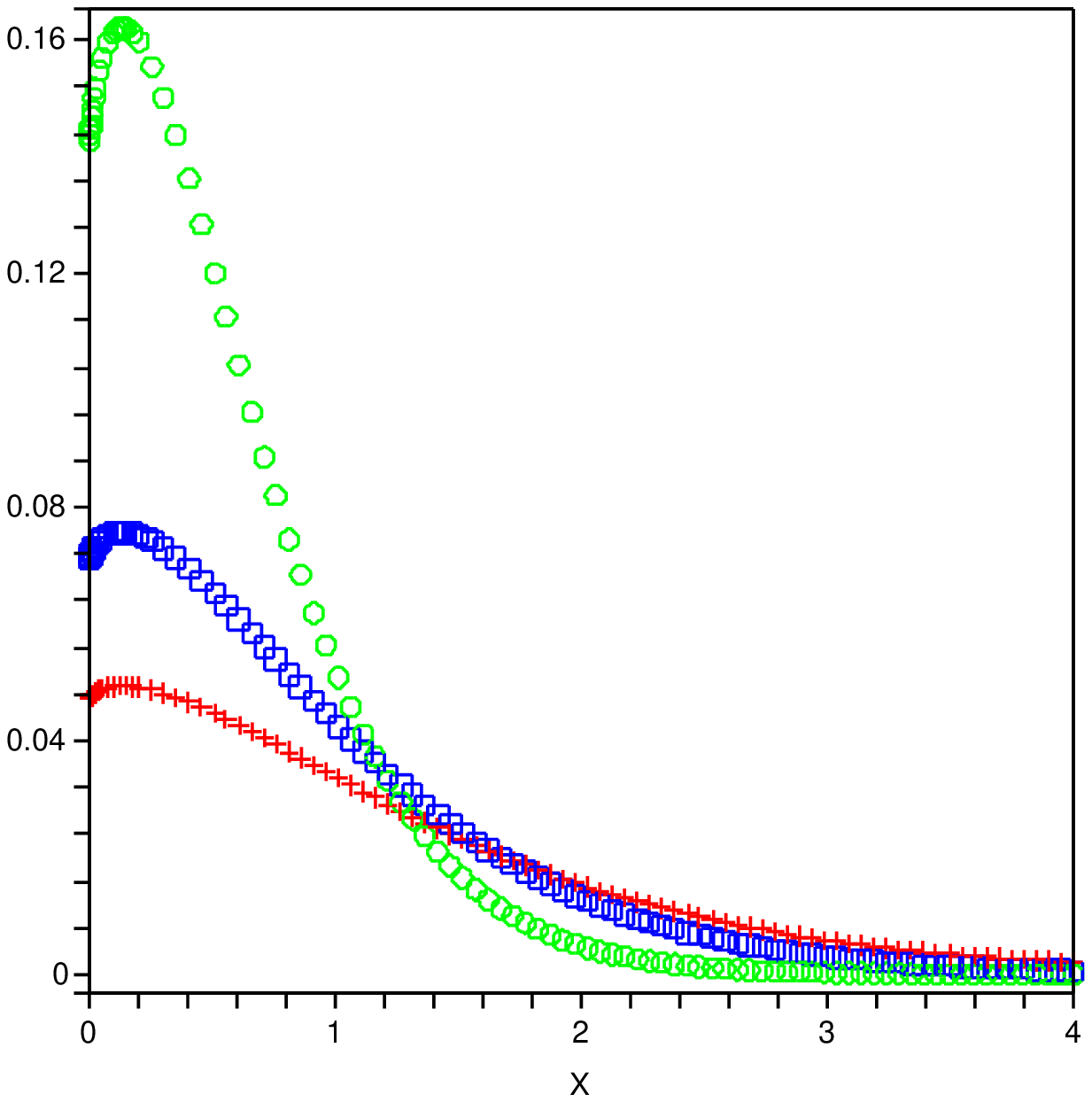}%
\\
Fig.1 $m\overline{\rho}(x)$ for $N=1(\bigcirc),2(\square),3(\times)$ in unit
of $c.$%
\end{center}}}%
{\parbox[b]{3.2214in}{\begin{center}
\includegraphics[
trim=0.000000in 0.000000in 0.053341in 0.053341in,
height=3.2214in,
width=3.2214in
]%
{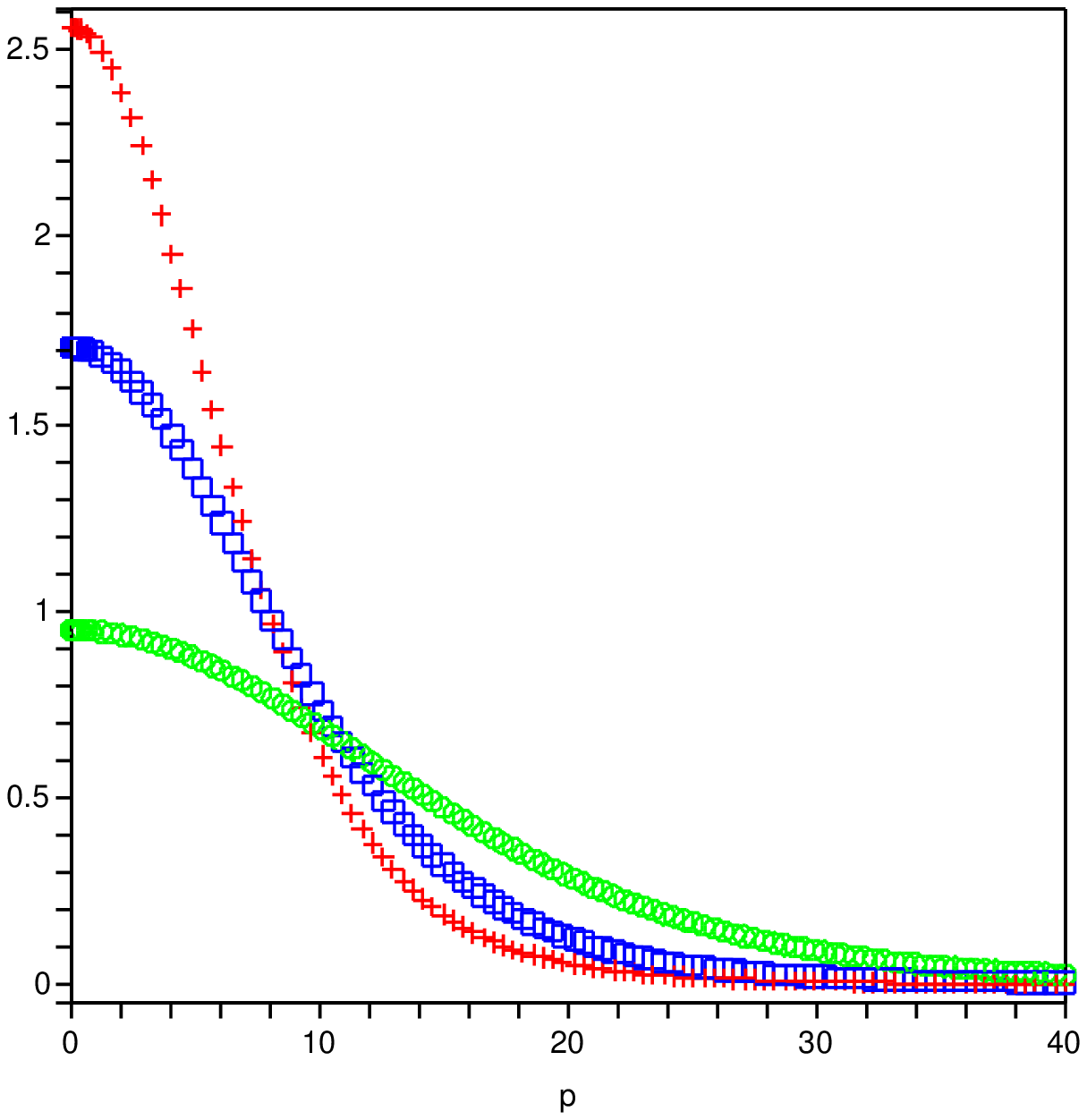}%
\\
Fig.2 $m\rho(p)$ for $N=1(\bigcirc),2(\square),3(\times)$ in unit of
$1/c,p=.1n.$%
\end{center}}}%

\section{Renormalization constant and order parameter}

In this section we consider the renormalization constant and bare mass in our
model.It is easy to evaluate the renormalization constant and bare mass
defined by the renormalization transformation
\begin{align}
\psi_{0}  &  =\sqrt{Z_{2}}\psi_{r},\overline{\psi}_{0}=\sqrt{Z_{2}}%
\overline{\psi}_{r},\nonumber\\
S_{F}^{0}  &  =Z_{2}S_{F},\frac{Z_{2}^{-1}}{\gamma\cdot p-m_{0}}=S_{F}(p)=\int
ds\frac{\gamma\cdot p\rho_{1}(s)+m\rho_{2}(s)}{p^{2}-s}.
\end{align}%
\begin{align}
Z_{2}^{-1}  &  =\int\rho_{1}(s)ds=\lim_{p\rightarrow\infty}\frac{1}%
{4}tr(\gamma\cdot pS_{F}(p))\nonumber\\
&  =\Gamma(D+1)c^{D}\lim_{p\rightarrow\infty}\int_{0}^{\infty}F(s)ds\frac
{\sqrt{-p^{2}}\sin((D+1)\arctan(\sqrt{-p^{2}}/(m-s))}{(p^{2}+(m-s)^{2}%
)^{(1+D)/2}}\nonumber\\
&  \rightarrow\sin(\frac{(D+1)\pi}{2})c^{D}\lim_{p\rightarrow\infty}\frac
{1}{\sqrt{-p^{2}}^{D}}=0.
\end{align}%
\begin{equation}
m_{0}Z_{2}^{-1}=m\int\rho_{2}(s)ds=\lim_{p\rightarrow\infty}\frac{1}%
{4}tr(p^{2}S_{F}(p))\rightarrow0.
\end{equation}
There is no pole and it shows the confinement for $D>0$.Order parameter
$\left\langle \overline{\psi}\psi\right\rangle $ is given as the integral of
the scalar part of the propagator in momentum space
\begin{equation}
\left\langle \overline{\psi}\psi\right\rangle =-TrS_{F}(x),
\end{equation}
In position space we evaluate directly from (40)%
\begin{align}
\left\langle \overline{\psi}\psi\right\rangle  &  =4\lim_{x\rightarrow
0}m\overline{\rho}(x)\nonumber\\
&  =-\frac{4mc^{D}}{4\pi}\exp(\frac{\gamma e^{2}}{8m})\lim_{x\rightarrow
0}x^{D-1}.
\end{align}
From the above equation we see that $\left\langle \overline{\psi}%
\psi\right\rangle $ $\neq0$ and finte only if $D=c/Nm=1.$In this case we have%
\begin{equation}
\left\langle \overline{\psi}\psi\right\rangle =-\frac{c^{2}}{\pi N}\exp
(\gamma)=-\frac{c^{2}}{\pi N}\cdot1.7810..
\end{equation}
with the relation between physical mass and coupling constant
\begin{equation}
m=\frac{c}{N}=\frac{e^{2}}{8}.
\end{equation}
The $N$ dependence of $\left\langle \overline{\psi}\psi\right\rangle $ is weak
at large $N$ in comparison with Dyson-Schwinger equation where the order
parameter vanishes fast with $N$[6,7,8,9,10,12].It is not difficult to test
the convergence for different values of $D$ numerically in momentum space%
\begin{align}
\left\langle \overline{\psi}\psi\right\rangle  &  =-4m\exp(\frac{\gamma e^{2}%
}{8m})\int_{0}^{\infty}\frac{p^{2}d\sqrt{-p^{2}}}{2\pi^{2}}\frac
{\Gamma(D+1)c^{D}}{2\sqrt{-p^{2}}}\nonumber\\
&  \times\int_{0}^{\infty}dsF(s)\frac{\sin((D+1)\arctan(\sqrt{-p^{2}}%
/(m-s))}{\sqrt{((m-s)^{2}+p^{2})^{D+1}}},(D=c/Nm).
\end{align}
At $D=1$ we derive directly (55) provided%
\begin{equation}
\int_{0}^{\infty}\frac{p^{2}dp}{(p^{2}+m^{2})^{2}}=\frac{\pi}{4m},\int
_{0}^{\infty}F(s)ds=1.
\end{equation}
We see vacuum expectation value is finite for $D=1$.This condition is
independent of the bare mass.However we cannot apriori determine the value
$D,$since $m$ is a physical mass$.m$ and zero momentum mass $\Sigma(0)$ is not
the same quantity in general but we assume they have the same order of
maginitude.In our model $\Sigma(0)$ can be evaluated as the value of the
inverse of the scalar part of the propagator at $p=0$ with $m_{0}=0.$In
numerical analysis of Dyson-Schwinger equation $\Sigma(0)$ damps fast with $N$
and is seen to vanish at $N=3$[6,7,8,9,10,12]$.$In our approximation if we set
$m$ equals to the value $B$ in (41) we obtain
\begin{equation}
D=\frac{2}{3-2\gamma}\simeq1.08
\end{equation}
which is very close to$1$.For large $N,D$ remains $O(1)$ but $C$ is $O(c/N)$
and mass generation is suppressed.For $D=1$ we have a similar solution of the
propagator at short distance which is known by the analysis of D-S.And we find
the $m$ and $\Sigma(0)$ are the same order.In the analysis of Gap
equation,zero momentum mass $\Sigma(0)$ and a small critical number of
flavours have been shown[6,7,8,9],in which same approximation was done because
the vacuum polarization governs the photon propagator at low energy.The
critical number of flavour $N_{c}$ is a consequence of the approximation for
the infrared dynamics as in quenched QED$_{4}$ where ultraviolet rigion is non
trivial and is not easy to find numerically.For fixed value of the coupling we
solved the coupled Dyson-Schwinger equation numerically and found that the
$\Sigma(0)$ which is,$O(e^{2}/4\pi)$ at $N=1$,the same order of magnitude as
we get in the quenched Landau gauge[11,12].The value of $\left\langle
\overline{\psi}\psi\right\rangle $ at $N=1$ is the same order of magnitude
$10^{-3}$ which is shown in[9].In the case of vanishing bare mass $m_{0}%
=0,$Ward-Takahashi-identity for axial vector currents%
\begin{equation}
\lim_{p\rightarrow q}(p-q)_{\mu}S_{F}(p)\Gamma_{5\mu}(p,q)S_{F}(q)=\{\gamma
_{5},S_{F}(p)\}
\end{equation}
implies exisitence of a massless pole in the $\Gamma_{5\mu}(p,q):F_{\pi}%
\chi(p-q)(p-q)_{\mu}/(p-q)^{2}$where $\chi$ is a B-S amplitude for psedoscalar
bound state.This is related to the propagator%
\begin{align}
F_{\pi}\chi(p)  &  =2m\rho(p)\gamma_{5},\nonumber\\
F_{\pi}\chi(x)  &  =2m\overline{\rho}(x)\gamma_{5}.
\end{align}
in our approximation.The scalar density $\overline{\psi}\psi(x)$ is understood
as the density of electron.$\nabla\overline{\psi}\psi(x)$ gives the gradient
energy of superfliud current.In our approximation or Dyson-Schwinger equation
we can determine the propagator and the density of electron at the same time
in the condensed phase.\qquad

\section{Summary}

We evaluate the fermion propagator in three dimensional QED with dressed
photon by the method of spectral function.In the evaluation of lowest order
matrix element for fermion spectral function we obtain finite mass
shift,Coulomb energy and gauge invariant position dependent mass which has the
same property in the the analysis of D-S equation.However there remains
infrared divegences such as linear and logarithmic ones which were
\ regulaized by an infrared cut-off $\mu.$Including vacuum polarization we
find that infrared behaviour is modified which is shown in (37)-(41) and we
can avoid the infrared divergences away from threshold.This result is
consistent with other analysis with dimensional regularization[4].For $c/Nm=1$
order parameter $\left\langle \overline{\psi}\psi\right\rangle $ is
finite,where we have the familiar relation of the mass and coupling constant
in $1/N$ approximation.In our analysis Coulomb energy at short distance is
logarithmically divergent and plays the same role of mass singularity in
four-dimension.This sets the renormalization constant $Z_{2}^{-1}=0$ for
arbitrary positive value of the coupling constant.If we assume the magnitude
of physical mass $m$ is $O(e^{2})$ in the Landau gauge at $N=1$,our results is
consistent with numerical analysis of coupled Dyson-Schwinger equation excepts
for wave renormalization [9,12].We find the similar structure of the
propagator in whole region except for $N$ dependence.The advantage of our
approximation is to get the super-fluid density $\overline{\psi}\psi(x)$ as
well as B-S amplitude for psedoscalar $\chi(x)$ by Ward-Takahashi-identity$.$

\section{Acknowledgement}

The author would like thank the members of CSSM for their hospitality during
his stay 2005.

\section{References\newline}

\noindent\lbrack1]R.Jackiw,S.Templeton,How Super-renormalizable interaction
cure their infrared divergences, Phys.Rev.D.\textbf{23}(1981)2291.\newline%
[2]R.Jackiw,L.Soloviev,Low-energy theorem approach to single-particle
singularities in the presence of massless bosons,Phys.Rev.\textbf{137}%
.3(1968)1485.\newline[3]Yuichi Hoshino,Mass singularity and confining property
in QED3,\textbf{JHEP0409}:048,2004.\newline%
[4]A.B.Waites,R.Delbourgo,Nonpertubativebehaviour in three-dimensional
QED,Int.J.Mod.Phys.\textbf{A7}(1992)6857.\newline%
[5]S.Deser,R.Jackiw,S.Templeton,Topologicallymassivegaugetheory,Ann,Phys.(NY)\textbf{140}%
(1982)372.\newline[6]T.Appelquist,D.Nash,Critical Behaviour
in(2+1)-DimensionalQED,Phys.Rev.Lett.\textbf{60}(1988)2575.\newline%
[7]E.Dagot,J.B.Kogut,and A.Kocic,A comupter simulation of chiral symmetry
breaking in(2+1)-dimensional QED with $N$ flavors.Phys.Rev.Lett\textbf{62}%
((1989)1083.\newline[8]T.Appelquist,L.C.R.Wijewarhana,PhaseStructureof
Non-CompactQED3and the Abelian Higgs Model, [arXiv:hep-ph/0403250].\newline%
[9]C.S.Fischer,R.ALkofer,T.Dahm,and
P.Maris,DynamicalChiralSymmetryBreakinginUnquenchedQED$_{3}$,\newline
Phys.Rev.D70073667(2004).\newline%
[10]S.J.Hands,J.B.Kogut,L.Scorzato,C.G.Strouthos,Phys.Rev.B70(2004)104501.\newline%
[11]Y.Hoshino,T.Matsuyama,Dynamical parity violation in QED in
three-dimensions with a two component massless fermion,Phys.lett.\textbf{B}
\textbf{222}(1989)493.\newline%
[12]Yuichi.Hoshino,Toyoki.Matsuyama,andChikage.Habe,FermionMassGenerationinQED$_{3}%
$,in theProceedings of the
WorkshoponDynamicalSymmetryBreaking,Nagoya,1989,0227-232.\newline%
[13]I.Mitra,R.Ratabole,H.Sharatchandra,Instability in scalar channel of
fermion-antifermion scattering amplitude in massless $QED_{3}$ and anomalous
dimension of composite operators[arXiv:hep-th/0601058].\newline[14]
L.F.Herbut,phys.Rev.\textbf{B66}%
,094504(2002)[arXiv:cond-matt/0202491];Phys.Rev.Lett.\textbf{88}%
(2002)047006.\newline[15]M.Franz,Z.Tesanovic,O.Vafee,Phys.Rev.B66(2002)05405[ArXivcond-matt/0110253];

M.Franz,Z.Tesanovic,Phys.Rev.Lett.87(2001)2291.\newline%
[16]K.Nishijima,\textbf{Fields and Particles},W.A.BENJAMIN,INC(1969).\newline%
[17]C.Itzykson,J.B.Zuber,Quantum field theory,McGRAW-HILL.\newline%
[18]L.S.Brown,Quantum field theory,Cambridege University Press(1992).\newline

\section{Appendices}

\subsection{Evaluation of the one-photon matrix element}

In this section we evaluate the one-photon matrix element $F$ in
(23),(25).Following the parameter tric%
\begin{align}
\frac{1}{k\cdot r} &  =i\lim_{\epsilon\rightarrow0}\int_{0}^{\infty}%
d\alpha\exp(i\alpha(k+i\epsilon)\cdot r),\nonumber\\
\frac{1}{(k\cdot r)^{2}} &  =\lim_{\epsilon\rightarrow0}\int_{0}^{\infty
}\alpha d\alpha\exp(i\alpha(k+i\epsilon)\cdot r),
\end{align}
we obtain the formulea to evaluate three terms in%
\begin{equation}
F=-e^{2}(\frac{\gamma\cdot r}{m}+1)\int\frac{d^{3}k}{(2\pi)^{2}}\exp(ik\cdot
x)\theta(k^{0})\delta(k^{2})[\frac{m^{2}}{(r\cdot k)^{2}}+\frac{1}{(r\cdot
k)}+\frac{d-1}{k^{2}}].
\end{equation}
First two terms are written explicitly by the photon ropagator $D_{F}(x)$%
\begin{equation}
F_{1}=-\int\frac{d^{3}k}{(2\pi)^{3}}\exp(ik\cdot x)D_{F}(k)\frac{1}{(k\cdot
r)^{2}}=-\lim_{\mu\rightarrow0}\int_{0}^{\infty}\alpha d\alpha D_{F}(x+\alpha
r)\exp(-\mu\alpha r).
\end{equation}%
\begin{equation}
F_{2}=-\int\frac{d^{3}k}{(2\pi)^{3}}\exp(ik\cdot x)D_{F}(k)\frac{1}{k\cdot
r}=-i\lim_{\mu\rightarrow0}\int_{0}^{\infty}d\alpha D_{F}(x+\alpha r)\exp
(-\mu\alpha r).
\end{equation}
Soft photon divergence corresponds to the large $\alpha$ region and $\mu$ is
an infrared cut-off$.$It is simple to evaluate the last term in $F$ by definition%

\begin{equation}
F_{L}=\int\frac{d^{3}k}{i(2\pi)^{3}}\exp(ik\cdot x)D_{F}(k)\frac{1}{k^{2}},
\end{equation}%
\begin{equation}
\frac{1}{4\pi^{2}}\int_{0}^{\infty}d\sqrt{k^{2}}\frac{\sin(\sqrt{k^{2}%
}\left\vert x\right\vert )}{\sqrt{k^{2}}\left\vert x\right\vert (k^{2}+\mu
^{2})}=\frac{1-\exp(-\mu\left\vert x\right\vert )}{8\pi\mu^{2}\left\vert
x\right\vert }.
\end{equation}
We have%
\begin{equation}
F=-ie^{2}m^{2}\int_{0}^{\infty}\alpha d\alpha D_{F}(x+\alpha r)-e^{2}\int
_{0}^{\infty}d\alpha D_{F}(x+\alpha r)+e^{2}(d-1)\int\frac{d^{3}k}{i(2\pi
)^{3}}\exp(ik\cdot x)D_{F}(k)\frac{1}{k^{2}}.
\end{equation}
In quenced case the above formulea for the evaluation of three terms in $F$
provided the position space propagator with bare mass
\begin{align}
D_{F}^{(0)}(x)_{+}  &  =\int\frac{d^{3}k}{i(2\pi)^{2}}\delta(k^{2}-\mu
^{2})\theta(k^{0})\exp(ik\cdot x)\\
&  =\frac{1}{i(2\pi)^{2}}\int_{0}^{\infty}\frac{\pi\sqrt{k^{2}}d\sqrt{k^{2}%
}J_{0}(\sqrt{k^{2}}\left\vert x\right\vert )}{2\sqrt{k^{2}+\mu^{2}}}%
=\frac{\exp(-\mu\left\vert x\right\vert )}{8\pi i\left\vert x\right\vert }.
\end{align}

\begin{align}
F  &  =-\frac{e^{2}}{8\pi}(\frac{\exp(-\mu\left\vert x\right\vert )}{\mu
}-\left\vert x\right\vert \operatorname{Ei}(\mu\left\vert x\right\vert
))-\frac{e^{2}}{8\pi\sqrt{r^{2}}}\operatorname{Ei}(\mu\left\vert x\right\vert
)+(d-1)\frac{e^{2}}{8\pi\mu^{2}\left\vert x\right\vert }(1-\exp(-\mu\left\vert
x\right\vert )),\nonumber\\
r^{2}  &  =m^{2},
\end{align}
where
\begin{align}
\operatorname{Ei}(z)  &  =\int_{1}^{\infty}\frac{\exp(-zt)}{t}dt,\\
\operatorname{Ei}(\mu\left\vert x\right\vert )  &  =-\gamma-\ln(\mu\left\vert
x\right\vert )+(\mu\left\vert x\right\vert )+O(\mu^{2}).
\end{align}
For the leading order in $\mu$ we obtain%
\begin{align}
-e^{2}F_{1}  &  =\frac{e^{2}}{8\pi}(-\frac{1}{\mu}+\left\vert x\right\vert
(1-\ln(\mu\left\vert x\right\vert -\gamma))+O(\mu),\\
-e^{2}F_{2}  &  =\frac{e^{2}}{8\pi m}(\ln(\mu\left\vert x\right\vert
)+\gamma)+O(\mu),\\
(d-1)e^{2}F_{L}  &  =\frac{e^{2}}{8\pi}(\frac{1}{\mu}-\frac{\left\vert
x\right\vert }{2})(d-1)+O(\mu).
\end{align}
Since we used the photon propagator with bare mass $\mu$ as $\exp
(-\mu\left\vert x\right\vert ),\left\vert F\right\vert $ falls fast and we
have a short distance contribution of \ $F$ which is negative in the Landau gauge.

\subsection{Analytic continuation of $S_{F}(p)$}

Here we show the analytic form of the quenched fermion propagator without mass
changing effects in Minkowski space by using the formulea%
\begin{align}
\arctan\text{h}(z)  &  =\frac{1}{2}\ln(\frac{1+z}{1-z})=-i\arctan(iz),(0\leq
z^{2}\leq1)\\
\arctan\text{h}(z)  &  =\frac{1}{2}\ln(\frac{1+z}{z-1})\pm\frac{i\pi}{2},\\
\text{arccoth}(z)  &  =\frac{1}{2}\ln(\frac{1+z}{z-1}),(1\leq z^{2}).
\end{align}
Principal part of the propagator in Minkowski space is continued to%
\begin{align}
S_{F}(p)  &  =\frac{(\gamma\cdot p+m)c^{D}\Gamma(D+1)\sinh((D+1)\text{arctanh}%
(\sqrt{p^{2}}/m)}{\sqrt{p^{2}}(m^{2}-p^{2})^{(D+1)/2}},(\left\vert \sqrt
{p^{2}}/m\right\vert \leq1)\\
&  =\frac{(\gamma\cdot p+m)c^{D}\Gamma(D+1)\sinh((D+1)\text{arccoth}%
(\sqrt{p^{2}}/m)}{\sqrt{p^{2}}(p^{2}-m^{2})^{(D+1)/2}},(\left\vert \sqrt
{p^{2}}/m\right\vert \geq1.\nonumber
\end{align}
The spectral function is a discontinuity in the upper half-plane of
$z^{2}=p^{2}/m^{2}\geq1$
\begin{equation}
\pi\rho(z)=-\frac{c^{D}}{2}\operatorname{Im}\frac{\Gamma(D+1)}{m\left\vert
z\right\vert (m^{2}-m^{2}z^{2})^{^{(D+1)/2}}}[(\frac{1+z}{1-z})^{(D+1)/2}%
-(\frac{1-z}{1+z})^{(D+1)/2}].
\end{equation}
For $z^{2}\geq1$%
\begin{align}
(\frac{1}{1-z^{2}})^{(D+1)/2}  &  =\left\vert \frac{1}{1-z^{2}}\right\vert
^{D+1}\exp(-(D+1)\pi i/2),\\
(\frac{1+z}{1-z})^{(D+1)/2}  &  =\left\vert \frac{1+z}{1-z}\right\vert
^{(D+1)/2}\exp(-(D+1)\pi i/2),z\geq1\\
(\frac{1-z}{1+z})^{(D+1)/2}  &  =\left\vert \frac{1-z}{1+z}\right\vert
^{(D+1)/2}\exp(-(D+1)\pi i/2),z\leq-1
\end{align}
we have%
\begin{equation}
\pi\rho(z)=\frac{c^{D}\Gamma(D+1)\sin((D+1)\pi)}{2m\left\vert z\right\vert
m^{D+1}}[(\frac{1}{z-1})^{D+1}\theta(z-1)-\left\vert \frac{1}{z+1}\right\vert
^{D+1}\theta(-(z+1)].
\end{equation}
The vanishment of the renormalization constant
\begin{equation}
Z_{2}^{-1}=\int zdz\rho(z)=0
\end{equation}
which is evaluated as the high energy behaviour of $S_{F}(p)$ (49) is seen by%
\begin{equation}
I=\int_{1+\epsilon}^{\infty}dz(\frac{1}{z-1})^{D+1}+\int_{-\infty
}^{-1-\epsilon}dz\left\vert \frac{1}{z+1}\right\vert ^{D+1}=0,
\end{equation}
where $\epsilon=\mu/m$ .

\end{document}